# LRSA: A new computational method for analyzing time course microarray data

Wei Wu[*], Nilesh B. Dave, and Naftali Kaminski

Dorothy P. and Richard P. Simmons Center for Interstitial Lung Disease, Division of Pulmonary, Allergy and Critical Care Medicine, University of Pittsburgh, School of Medicine, Pittsburgh, PA 15213, USA.

## ABSTRACT

**Motivation:** Time course data obtained from biological samples subject to specific treatments can be very useful for revealing complex and novel biological phenomena. Although an increasing number of time course microarray datasets becomes available, most of them contain few biological replicates and time points. So far there are few computational methods that can effectively reveal differentially expressed genes and their patterns in such data.

**Results:** We have proposed a new two-step nonparametric statistical procedure, LRSA, to reveal differentially expressed genes and their expression trends in temporal microarray data. We have also employed external controls as a surrogate to estimate false discovery rates $FDR_{EC}$ and thus to guide the discovery of differentially expressed genes. Our results showed that LRSA reveals substantially more differentially expressed genes and have much lower $FDR_{EC}$ than two other methods, STEM and ANOVA, in both real data and the simulated data. Our computational results are confirmed using real-time PCRs.

**Contact:** wuw2@upmc.edu

# INTRODUCTION

Time course data can be extremely useful for revealing complex and novel biological phenomena underlying normal physiological conditions, developmental stages, and various diseases in animals. As microarray technology becomes indispensable for biological research, there is an increasing number of time course microarray datasets generated and becoming available to the public. However, due to the high cost of microarrays, these datasets are limited in that they contain a small number of time points and sample few biological replicates for each time point (Ernst, et al., 2005). As a result, conventional computational techniques for analyzing time course data cannot be applied to these data, e.g., Hidden Markov Models (HMM) and other dynamic models, wavelets and other adaptive methods (Daubechies, 1992; Hardle, et al., 1997); these methods require plenty of data to ensure accurate estimate of structures in the data (Loader, 1999).

Approaches for analyzing time course microarray data usually focus on two tasks, one is to identify differentially expressed (DE) genes among different classes of samples − this task can also be considered as a gene-filtering step, and the other is to cluster these genes into different groups. As a result, time course approaches can also be classified into two types: i) the methods which were designed mainly to characterize significant DE genes and then to cluster the genes using standard clustering algorithms (e.g., hierarchical clustering and K-means); and ii) the methods designed mainly for clustering genes which were usually passed through simple gene filters.

Methods focusing on the identification of significant DE genes include linear-regression based methods (Conesa, et al., 2006; Xu, et al., 2002). These methods are limited in that i) they are parametric; since gene expression patterns are usually unknown, the models used in the methods may not capture expression trends of all genes. ii) these methods assume that gene expression levels are homoscedastic (i.e., constant variance) over time. This assumption can be oversimplified, since gene expression patterns in response to a specific treatment (e.g., hypoxia) can be drastically different at different time points. Other methods include spline-based methods (Bar-Joseph, et al., 2003; De Hoon, et al., 2002; Storey, et al., 2005); they are nonparametric and can capture flexible patterns in time course data. However, these approaches can be overfitting for short time course data (Ernst, et al., 2005); besides, they also assume data is homoscedastic. In addition, the t- and F-statistics were calculated in some of these methods to determine significant DE genes (De Hoon, et al., 2002; Storey, et al., 2005); although these statistics are valid for static data, they may not be stringent enough for temporal data (see Discussion).

Early clustering-focused methods employed hierarchical clustering and principle component-related analyses to cluster genes (Alter, et al., 2000; Eisen, et al., 1998). The drawbacks of these approaches are that they treated time course data as static data, and did not account for temporal structures in the data. A few methods proposed later accounted for the dynamic nature of the data (Ramoni, et al., 2002; Schliep, et al., 2003), but they also suffer from the overfitting problem as pointed out in (Ernst, et al., 2005). Other approaches have proposed to build computational models using predefined profiles that required domain knowledge. The limitations of these methods include that i) genes expression profiles under certain biological conditions are often unknown; ii) the methods may lead to the enrichment of the artifacts which resemble the predefined patterns; and iii) genes whose expression profiles are not predefined will be ignored. To improve over these methods, strategies that assigned genes into different clusters using predefined changes of expression patterns (e.g., STEM) have been proposed (Ernst and Bar-Joseph, 2006; Kim and Kim, 2007). Although prior knowledge is not required for these methods, defining changes of expression patterns can be rather arbitrary.

Multiple testing problems are prevalent in microarray analysis. Since there are usually tens of thousands of genes on the array, detecting false positive genes can be increased by chance. Many procedures for controlling multiple testing have been proposed to control the number of false positives or false discovery rates (FDRs) (Benjamini and Hochberg, 1995; Dudoit, et al., 2003). However, the procedures can become overly conservative due to the huge number of genes on the arrays, that is, although they control false positives or FDRs well, the power of detecting true positives is often low.

In this work, we developed a new two-step nonparametric procedure, LRSA (*Local Regression and Spectral Analysis*), to analyze time course microarray data. In the first step, LRSA detects significant DE genes at different time points with respect to the control time point using a local smoothing procedure. To characterize temporal DE genes accurately, LRSA estimates simultaneous confidence intervals using a method which accounts for heteroscedasticity in the data. In addition, we used external controls as a surrogate to estimate FDRs and thus to guide the detection of DE genes. We show that LRSA can detect significant DE genes effectively and with high power. In the second step, LRSA clusters the DE genes resulted from the first step using a spectral clustering algorithm. Finally, we show that our computational results are verified using experimental approaches.

## DATA
*Real Data*: The data were collected to test the difference in responses between a control group of rats exposed to hypoxia for 0 day and the test groups of rats exposed to hypoxia for either 1, 3, 7, 14 or 30 days. In our initial experimental design, each group contained 3 biological replicates (i.e., microarray data were obtained from RNAs extracted from 3 different rats). However, due to the quality issues, some microarrays were eliminated before data pre-processing, and therefore some test groups (i.e., 3 and 30 days) contain fewer than 3 biological replicates. To monitor the consistency of microarray data collected from different batches of experiments, RNAs from some rats were used for multiple times (in different batches) to generate microarray data, therefore there are multiple technical replicates for some biological replicates (see Table 1s in Supplementary Materials for details). The dataset contains 30 arrays in total. The arrays used in this dataset are CodeLink UniSet Rat I Bioarrays containing pre-validated oligonucleotide probes targeting about 10K transcripts in the rat genome. Our protocols for performing microarray experiments and pre-processing microarray data are described in detail in Supplementary Materials.

*Simulated Data*: To generate simulated data, we first randomly selected one technical replicate (if multiple replicates were available) for each biological replicate for each time point. Then for time points 1, 3, 7, and 30 days, we randomly selected one biological replicate for each of these time points. For time points 0 and 14 days, we kept all 3 biological replicates to ensure that simultaneous confidence intervals for the simulated data can be estimated reasonably well (see METHODS for details).

## METHODS
*Local Regression and Spectral Analysis (LRSA)*: We have developed a flexible yet rigorous nonparametric statistical procedure, LRSA, which can detect genes differentially expressed at different time points with respect to the controls (0 day), and identify and cluster temporal expression patterns of these genes. The details of the LRSA procedure are described as follow.

*Fitting time course expression data using a nonparametric local regression smoothing model*: First, temporal expression data for each transcript is fitted using a local polynomial regression model. In particular, if we let *t* denote a vector of the time points corresponding to the lengths of the hypoxia treatment for all arrays in the hypoxia dataset, and let $s_i(t)$ denote a vector of the normalized intensity values (i.e., signals) for gene *i* corresponding to *t* in the data, then the nonparametric local regression model can be written as:

$$s_i(t) = f_i(t) + \varepsilon_i(t), \qquad (1)$$

where $f_i(t)$ is a smoothing function, and can be estimated by fitting a polynomial quadratic model within a smoothing parameter or bandwidth $h_i$; $\varepsilon_i(t)$ are errors normally distributed with mean 0 and variance $\sigma_i^2(t)$.

*Choosing the optimal bandwidth*: We used the locally weighted sums of squares (Sun and Loader, 1994; Wasserman, 2006) to select an optimal smoothing function at a particular fitting point $t_k$. The bandwidth $h_i$ is selected optimally using the generalized cross validation (GCV) score (Sun and Loader, 1994), which is an approximation of the leave-one-out cross-validation score. In this work, the local regression models were estimated using implementation in the R package/function *locfit/locfit($S_i$ ~ lp(t, deg=2, nn=$h_i$), data=mydata)*; and the GCV scores were calculated using the implementation of the R package/function *locfit/gcv(local regression model)["gcv"]* (Loader, 2006).

*Calculating simultaneous confidence intervals*: For estimating significant DE genes over time with respect to the controls, we used a method due to Faraway and Sun (1995) to calculate an approximate simultaneous $1-\alpha$ (95%) confidence interval $I_i(t)$ for the expected (or fitted) intensity values at time points *t* (denoted as $\hat{f}_i(t)$). This method is described in detail in (Faraway and Sun, 1995; Wasserman, 2006), and

$$I_i(t) = \left(\hat{f}_i(t) \pm c_i \hat{\sigma}_i(t) \|l(t)\|\right) \qquad (2)$$

for some $c > 0$, and $l(t)$ satisfies:

$$\hat{f}_i(t) = l(t)^T s_i(t). \qquad (3)$$

In this work, we set the *p*-value for estimating significant DE genes to be 0.05. The simultaneous confidence bands were calculated using implementation in the R package/function *locfit/crit (local regression model, cov=(1 - 0.05))* (Loader, 2006).

Since in the simulated data, only one biological replicate is present for some time points (1, 3, 7 and 30 days), we estimated simultaneous confidence intervals for this data using a modified procedure described in detail in Supplementary Materials.

*Controlling for multiple testing*: To control for multiple testing in the above procedure, we used a Bonferroni-type procedure, that is, the simultaneous confidence intervals are estimated at the $1-(\alpha/m)$ level, where *m* represents the number of time points (6) or the number of genes (10014), depending on whether multiple testing is controlled for the time points or the genes (Wasserman, 2006).

Finally, we determined genes as significantly differentially expressed only when their expression at some time point *k* relative to the controls satisfies i) $p\text{-value} \leq 0.05$, and ii) *fold-change* $\geq 2$ (on the original data scale).

*Calculating FDRs for external controls*: CodeLink UniSet Human I Bioarrays contain 408 bacterial control probes (68 unique probes, each spotted 6 times) on each array. Of the 68 unique probes, 18 are positive control probes and 50 are negative control probes. Since each of these probes is present at the same location with the same concentration on each array, they can be considered as true nulls and should not be identified as differentially expressed. Here, we propose to employ external control probes to monitor the FDRs associated with the method under study. If we let $N_{EC,DE}$ denote the number of external control probes identified as differentially expressed by a method, and let $N_{T,DE}$ denote the total number of the DE genes identified by the same method, then following (Benjamini and Hochberg, 1995), a surrogate FDR, which we call FDR for external controls $FDR_{EC}$, can be defined as

$$FDR_{EC} = N_{EC,DE} / N_{T,DE} . \qquad (4)$$

*Spectral clustering*: In order to capture temporal expression trends among significant DE genes, we clustered these genes using their expression profiles with a spectral clustering algorithm. This method clusters data points using eigenvectors of matrices derived from the data (Ng, et al., 2002); it is nonparametric, and clusters data based on the connectivity rather than the concentration of the data points, therefore it is particularly suitable for clustering high-dimensional data. It has been shown that spectral clustering reveals more accurate, robust and meaningful patterns than other clustering algorithms, such as K-means (Ng, et al., 2002). In this work, we have implemented a spectral clustering algorithm in Matlab based on Ng, et. al. (2002). We employed correlation coefficients as the affinity matrix for the similarity measurement between genes, that is,

$$d(s_i, s_j) = \sum_{t=1}^{T} (s_i(t) - \bar{s}_i)(s_j(t) - \bar{s}_j) \bigg/ sqrt\left( \sum_{t=1}^{T} (s_i(t) - \bar{s}_i)^2 (s_j(t) - \bar{s}_j)^2 \right),$$

where $d(s_i, s_j)$ denotes the affinity matrix for genes *i* and *j*, and *t* is the time point for the fitted data and $0 \leq t \leq T$.

To cluster genes differentially expressed over time using the spectral algorithm, we have used the expected temporal intensity values $\hat{f}_i(t)$ for any gene *i* derived from the local regression smoothing model described above as inputs, rather than using its normalized intensity values $s_i(t)$. The reasons for doing so are: i) $\hat{f}_i(t)$ is presumably less noisy than $s_i(t)$; and ii) since $\hat{f}_i(t)$ can be calculated for any *t* satisfying $0 \leq t \leq 30$, clustering analysis can be performed using $\hat{f}_i(t)$ corresponding to many more time points than the actual experimental time points; as such, gene expression trends in the data can be captured more robustly. In this work, we have used $\hat{f}_i(t)$ corresponding to 31 time points (spreading evenly between 0 and 30 days) as inputs for clustering analysis. $\hat{f}_i(t)$ is standardized (with mean 0, and the standard deviation 1) before the data for all genes is clustered.

**Other Methods:** We compared LRSA with two other methods: STEM and ANOVA. STEM clusters genes in the following way: it first filters genes using simple criteria based on the minimum change of gene expression values, the number of missing values and correlation between replicates; then it assigns a fraction of the genes that are passed through the filter into a predefined set of clusters, which can be further merged based on the significant Gene Ontology annotation (Ernst and Bar-Joseph, 2006; Ernst, et al., 2005). The STEM software can be downloaded from http://www.cs.cmu.edu/~jernst/stem. We used the default setting in STEM for clustering analysis in this work.

We also compared LRSA with ANOVA, which we have previously used to detect genes differentially expressed over time in hypoxia. A quadratic regression model was used to fit the response (gene expression values) and the explanatory variables (days of the treatment) in ANOVA and the details of which can be found in (Wu, et al., 2005). The *p*-values of the F statistics were adjusted using the Benjamini & Hochberg FDR procedure (Benjamini and Hochberg, 1995).

***Quantitative real-time (RT) PCR***: The detailed protocol for RT-PCR can be found in (Pardo, et al., 2005).

## RESULTS

### 1.1 Detecting significant DE genes in real data.

#### *4.1.1 LRSA and estimating FDRs using external controls*

LRSA is a two-step nonparametric procedure developed for revealing significant DE genes and their patterns in time course microarray data. In the first step, LRSA calculates the expected temporal expression levels $\hat{f}_i(t)$ and its simultaneous 95% confidence intervals $I_i(t)$ for any gene *i*, then it determines whether gene *i* is a significant DE gene using $I_i(t)$. In addition, we propose to monitor the FDRs of LRSA using external control probes on the arrays. Since true DE genes are unknown in real data, it is impossible to calculate exact FDRs without using experimental approaches; instead we employed external control probes on the arrays as a surrogate to estimate $FDR_{EC}$. In the second step, LRSA clusters the DE genes resulted from the first step using a spectral clustering algorithm.

Figure 1 illustrates the 4 genes identified as differentially expressed by LRSA in the hypoxia data; it shows that LRSA can detect not only genes exhibiting a smooth, temporal expression curve with a single peak (Figure 1B-1D), but also genes exhibiting cyclic patterns (Figure 1A).

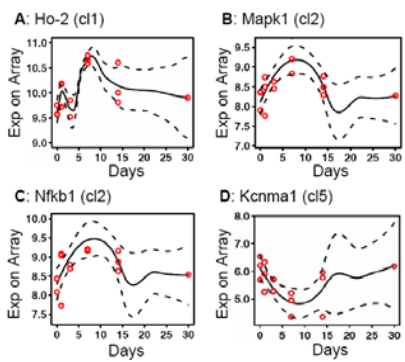

**Fig. 1.** Temporal expression patterns of the 4 significant DE genes identified by LRSA in the hypoxia data. (A) Gene expression pattern of the Ho-2 gene (residing in cluster 1) fitted by LRSA. Each red dot in the plot represents gene expression level (shown in the y axis) in each individual rat exposed to hypoxia for the designated days (shown in the x axis). The solid black curve shows the smoothing curve fitted to the microarray data by LRSA; the dotted curves shows the simultaneous 95% confidence intervals estimated for the fitted data. The other 3 genes shown are: Mapk1 (B), Nfkb1 (C), and Kcnma1 (D).

*4.1.2  Comparing LRSA with two other methods using real data*

Next we compared LRSA with two other methods: STEM and ANOVA, in terms of the ability to reveal significant DE genes.

Table 1 shows that LRSA revealed substantially more genes than STEM and ANOVA when the same multiple testing procedure was used. For example, without adjusting for multiple testing and using the 2-fold-change as a cutoff, LRSA revealed 1525 significant DE genes; whereas ANOVA revealed only 237 significant genes. $FDR_{EC}$ for both methods was 0.

STEM, however, employs a different procedure for revealing temporal expression patterns of informative genes (see METHODS for details). For the hypoxia data, STEM revealed that 2351 genes (not necessarily significant) passed the gene filter, among which 69 were external controls and thus $FDR_{EC} = 0.03$; then it assigned 492 genes (out of the 2351 genes, and can be considered as significant) into clusters, among which 33 were external controls, and thus $FDR_{EC} = 0.07$. Moreover, our results showed that indeed with STEM, external controls had significantly higher probability to be identified as differentially expressed than the other target transcripts on the arrays (Fisher's exact tests, odd ratio = 1.9, p-value = 8E-4; Table 1); whereas that the opposite was true with LRSA (e.g., without multiple testing control, odd ratios = 0 − 0.07, p-values = 5E-34 − 4E-28; Table 1). These results indicate that LRSA is more powerful than ANOVA and STEM, and yields significantly lower $FDR_{EC}$ than STEM.

We also compared LRSA with and without multiple testing control (Table 1). Not surprisingly, LRSA without multiple testing control had the most power (revealing 1525 genes with the 2-fold-change cutoff), whereas LRSA with multiple testing control for the number of genes (10014 genes on the array) had the least power (revealing 86 genes with the 2-fold-change cutoff). Unexpectedly, however, LRSA with and without multiple testing control had similar $FDR_{EC}$ (in the range of 0-0.004, Table 1). These results suggest that LRSA without multiple testing control can reveal substantially more DE genes and thus more powerful, yet without suffering from noticeably higher $FDR_{EC}$, than the same procedure with multiple testing control.

*4.1.3  Verification of significant DE genes*

In order to verify our findings revealed by LRSA, we did literature search as well as RT-PCRs. We found that there are 4 genes previously known to be transcriptionally regulated and play key roles in the rat lungs during hypoxia: the up-regulated Mapk1, Nfkb1, and Ho-2, and the down-regulated Kcnma1. Our results showed that LRSA without multiple testing control identified all the 4 genes correctly (Table 2).

Next, we selected 6 genes (gene1 − gene6) from the list of the significant DE genes identified by LRSA without multiple testing control, and verified their expression levels using RT-PCRs. These genes were also chosen to facilitate further verification of the clustering results by LRSA (see Section 4.3). Our results showed that the 6 selected genes were indeed differentially expressed using RT-PCR verification (data not shown). Table 2 summarizes the prediction accuracy of our evaluated methods. It shows that LRSA without multiple testing control (100% prediction accuracy, see the legend of Table 2 for details) performed substantially better than STEM (30% prediction accuracy) and ANOVA with and without multiple testing control (0% prediction accuracy). Table 2 also shows that LRSA without multiple testing control performed better than LRSA with multiple testing controlled for the number of time points (80% prediction accuracy), both of which performed significantly better than LRSA with multiple testing controlled for the number of genes (10% prediction accuracy). These results confirm that LRSA is more powerful than STEM and ANOVA, and suggest again that multiple testing control may be unnecessary and that $FDR_{EC}$ can be used to guard and assess the prediction accuracy of the statistical methods.

### 1.2 Comparing LRSA with STEM and ANOVA using simulated data.

Time course microarray datasets often contain few biological and technical replicates due to costly microarray experiments. To test how effective LRSA performs on such data, we did a simulation study. We generated a 'parsimonious' dataset using the strategy described in detail in the DATA (section 2), then we compared LRSA with STEM and ANOVA using the simulated data.

Table 3 shows that when the same multiple testing procedure was used, LRSA revealed substantially more significant DE genes, yet yielded significantly lower $FDR_{EC}$, than STEM and ANOVA. It is also noticeable that although LRSA without multiple testing control identified more DE genes than LRSA with multiple testing control, they had the same $FDR_{EC}$, suggesting again that multiple testing control is unnecessary for LRSA.

Since the simulated data was derived from the real hypoxia data, we calculated the percentage of the DE genes revealed in the hypoxia data which overlapped with the DE genes revealed in the simulated data using the same method. Table 3 shows that without multiple testing control, LRSA revealed 32-42% of genes differentially expressed in both the real and the simulated data, while that STEM revealed only 8% of such genes, and ANOVA 22-24% of such genes. These results suggest that LRSA is the most robust while STEM is the least robust of the 3 evaluated methods.

We also examined how effective LRSA, STEM, and ANOVA detected the 4 known DE genes and the 6 verified DE genes using the simulated data. Table 4 shows that LRSA without multiple testing control (40-60% prediction accuracy) performed significantly better than STEM (10% prediction accuracy) and ANOVA (0% prediction accuracy).

### 1.3 Clustering significant DE genes using LRSA in real data.

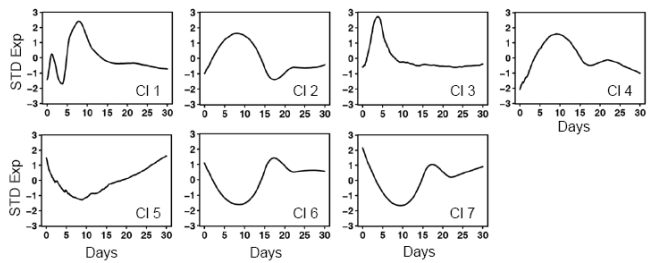

**Fig. 2.** The average gene expression pattern of each cluster. This pattern was obtained by taking the median expression value of all the genes for each time point in each cluster. The x-axis represents the lengths of the hypoxia treatment; the y-axis represents the standardized expression values of genes in each cluster (see METHODS for details).

Using LRSA (without multiple testing control and the 2-fold-change cutoff), we assigned 1525 significant DE genes in the hypoxia data into 7 distinct clusters. The average gene expression pattern of each cluster is displayed in Figure 2. In order to verify the clustering accuracy of LRSA, we further validated the temporal expression patterns of the 6 selected genes (gene1 – gene6) using RT-PCRs. These genes were selected because i) they resided in the clusters which exhibited distinct patterns: gene1 and gene2 (cl 1), gene3 (cl 2), gene4 (cl 3), gene5 (cl 4), and gene6 (cl 6) (see Figure 2 for the patterns of the clusters); and ii) the temporal expression patterns of these genes bore close resemblance to the average expression pattern of their respective cluster. Our results showed that overall the temporal expression patterns of the 6 genes validated using RT-PCRs agreed well with their patterns estimated by LRSA (Table 3s in Supplementary Materials).

We also examined clustering results revealed by STEM, which assigned 492 genes into 9 clusters. STEM only assigned 3 of the 6 validated genes (gene2-gene4) into clusters. Table 3s (Supplementary Materials) shows that compared to LRSA (Pearson correlation 0.79-0.95), temporal expression patterns of the 3 genes estimated by STEM correlated less well with those measured using RT-PCRs (Pearson correlation 0.60-0.93). These results indicate that LRSA identifies and clusters gene expression patterns more accurately than STEM.

## DISCUSSION

In this work, we proposed a new two-step nonparametric procedure LRSA to identify temporal DE genes and their expression patterns in time course microarray data. LRSA does not take any model assumption on the data, and uses a method due to Faraway and Sun (1995) to calculate approximate simultaneous 95% confident intervals to estimate the variances of gene expression values over different time points. These confident intervals have particularly desirable properties including: i) they are simultaneously valid for all the time points in the data. The t- and F-statistics have been used to determine significant genes in the time course data (De Hoon, et al., 2002; Storey, et al., 2005). However, these statistics are only valid for static data, but not for temporal data, because the confidence intervals derived from them cannot guarantee the simultaneous coverage of $\hat{f}(t)$ for the desired confidence for all the time points (Wasserman, 2006). ii) The Faraway and Sun method does not assume data is homoscedastic over time; it allows the estimate of heteroscedasticity in the data. Our results demonstrate that LRSA can indeed flexibly yet accurately reveal gene expression patterns of various shapes and various heteroscedasticity (Figures 1 and 2).

We have also proposed a novel metric $FDR_{EC}$ to monitor the FDRs using external control probes on the arrays. We showed that $FDR_{EC}$ can be used to guide the detection of significant DE genes. Our results indicate that LRSA can reveal significantly more genes and thus more powerful, yet with much lower $FDR_{EC}$, than STEM and ANOVA in both the real and the simulated data. Moreover, LRSA identifies substantially higher percentage of the overlapping significant DE genes in both the real and the simulated data, and hence it is a more robust method than STEM and ANOVA.

Classical multiple testing procedures, such as the Bonferroni procedure, aim to control the family-wise error rate (FWER) in a strong sense, that is, they control the probabilities of committing at least one Type I error at the level α under any combinations of the true and false null hypothesis. These procedures are known to be too conservative, especially when the number of the test hypothesis is large. To overcome this limitation, several strategies have been proposed. Among them, the procedures for controlling FDRs are less stringent yet more powerful than those for controlling FWER (Benjamini and Hochberg, 1995). Some went even further and recommended to avoid the use of the multiple testing procedures (Saville, 1990). The argument for doing so is that even though the probabilities of committing a Type I error can be reduced by multiple testing control, the probabilities of committing Type II errors can increase sharply. A concern for this recommendation is that an unguarded use of a statistical method can be problematic – it may cause a large number of false positives or high FDRs due to the intrinsic flawed design of the method.

Indeed our results showed that LRSA with the Bonferroni procedure is substantially less powerful yet yields the same $FDR_{EC}$ as LRSA without multiple testing control (Tables 1-4). Moreover, even when we applied ANOVA in conjunction with the less stringent FDR-controlling procedure to our data, we observed the similar results (Tables 1 and 3), suggesting that the FDR-controlling procedures are still too conservative given that there are a huge number of genes on the arrays. Our empirical findings suggest that we may avoid multiple testing control, instead use $FDR_{EC}$ as a surrogate to monitor the FDR of a statistical method, and thus safe guard the detection of the DE genes in the high-throughput microarray data. Indeed, we found that some unguarded methods (without multiple testing control) can lead to high FDR (Tables 1 and 3, "STEM"); nevertheless with the aid of $FDRs_{EC}$, we were able to assess different methods objectively, and select the best approach which is both powerful and yields low FDRs for further computational analysis.

We have also shown that LRSA can effectively and accurately reveal distinct gene expression patterns (Table 3s, Figure 2). Its potential limitations include: i) a sufficient number of time points are needed for the accurate estimate of the expected expression levels of genes; and ii) in order to estimate simultaneous confidence intervals correctly, multiple biological replicates are needed for at least a few time points. These limitations, however, are also common for other time course methods. Our future work aims to systematically compare LRSA with other time course approaches, and to recommend the best method to warrant accurate and robust temporal microarray analysis.

**FUNDING**
WW's work is funded by the NIH grant P50 HL084932 and the NSF grant CCF-0523757.

**Tables:**

**Table 1.** Comparison of LRSA with STEM and ANOVA using real data. This table shows the numbers of genes and external controls (EC) revealed as differentially expressed by each method. $FDR_{EC}$ is calculated as described in the METHODS. The odd ratios and p-values from Fisher's exact tests are shown to indicate whether a method has higher probability to identify ECs as differentially expressed than the real target transcripts on the array.

|  | Correction Factors* | Fold Change | No of Total DEs | No of ECs | $FDR_{EC}$ | Odd Ratio | p-value |
|---|---|---|---|---|---|---|---|
| LRSA | none | 1.5 | 2456 | 9 | 0.004 | 0.07 | 5E-34 |
| | none | 2.0 | 1525 | 0 | 0 | 0 | 4E-28 |
| | time points | 1.5 | 1136 | 4 | 0.004 | 0.08 | 3E-15 |
| | time points | 2.0 | 787 | 0 | 0 | 0 | 3E-14 |
| | genes | 1.5 | 115 | 0 | 0 | 0 | 0.01 |
| | genes | 2 | 86 | 0 | 0 | 0 | 0.04 |
| STEM | none | - | 492 | 33 | 0.07 | 1.9 | 8E-4 |
| ANOVA | none | 1.5 | 379 | 9 | 0.02 | 0.6 | 0.09 |
| | none | 2 | 237 | 0 | 0 | 0 | 1E-4 |
| | genes | 1.5 | 1 | 0 | 0 | 0 | 1.0 |
| | genes | 2 | 1 | 0 | 0 | 0 | 1.0 |

\* indicates whether the method for revealing significant DE genes was used in conjunction with a multiple testing procedure; if it was, whether the procedure was controlled for the number of the time points (6), or the number of the genes on the array (10014).

**Table 2.** Comparison of the prediction accuracy of LRSA, STEM and ANOVA using real data. This table shows the number of the DE genes (4 known and the 6 verified, see Section 4.1.3) identified respectively as differentially expressed by each method. The prediction accuracy is calculated as (the number of the 4 known genes identified by a method + the number of the 6 verified genes identified by the same method)/10*100%

|  | Correction Factors | Fold Change | Four Known Genes* | Six Verified Genes | Prediction Accuracy (%) |
|---|---|---|---|---|---|
| LRSA | none | 1.5 | 4 | 6 | 100 |
| | none | 2 | 4 | 6 | 100 |
| | time points | 1.5 | 4 | 4 | 80 |
| | time points | 2 | 4 | 4 | 80 |
| | genes | 1.5 | 1 | 0 | 10 |
| | genes | 2 | 1 | 0 | 10 |
| STEM | none | - | 1 | 3 | 40 |
| ANOVA | none | 1.5 | 0 | 0 | 0 |
| | none | 2 | 0 | 0 | 0 |
| | genes | 1.5 | 0 | 0 | 0 |
| | genes | 2 | 0 | 0 | 0 |

\* Four known DE genes: Ho-2, Mapk1, Nfkb1, Kcnma1.

**Table 3.** Comparison of LRSA with STEM and ANOVA using simulated data. The details of this table are described in the legend of Table 1.

|  | Correction Factors | Fold Change | No of Total DEs (% Ov*) | No of ECs | $FDR_{EC}$ | Odd Ratio | p-value |
|---|---|---|---|---|---|---|---|
| LRSA | none | 1.5 | 2277 (42) | 8 | 0.004 | 0.07 | 1E-31 |
| | none | 2 | 1184 (32) | 6 | 0.005 | 0.11 | 4E-14 |
| | time points | 1.5 | 1437 (38) | 5 | 0.003 | 0.08 | 2E-19 |
| | time points | 2 | 837 (35) | 4 | 0.005 | 0.11 | 3E-10 |
| | genes | 1.5 | 370 (24) | 2 | 0.005 | 0.14 | 7E-05 |
| | genes | 2 | 254 (27) | 1 | 0.004 | 0.1 | 6E-4 |
| STEM | none | - | 231 (8) | 18 | 0.08 | 2.2 | 3E-3 |
| ANOVA | none | 1.5 | 302 (24) | 7 | 0.02 | 0.60 | 0.1 |
| | none | 2 | 123 (22) | 1 | 0.008 | 0.21 | 0.05 |
| | genes | 1.5 | 8 (0) | 0 | 0 | 0 | 0.7 |
| | genes | 2 | 3 (0) | 0 | 0 | 0 | 0.9 |

\* represents the percentage of significant DE genes detected in real data which overlaps with the DE genes identified in the simulated data by the same method.

**Table 4.** Comparison of the prediction accuracy of LRSA, STEM and ANOVA using simulated data. The details of this table are described in the legend of Table 2.

|  | Correction Factors | Fold Change | Four Known Genes | Six Verified Genes | Prediction Accuracy (%) |
|---|---|---|---|---|---|
| LRSA | none | 1.5 | 2 | 4 | 60 |
| | none | 2 | 1 | 3 | 40 |
| | time points | 1.5 | 0 | 2 | 20 |
| | time points | 2 | 0 | 2 | 20 |
| | genes | 1.5 | 0 | 1 | 10 |
| | genes | 2 | 0 | 1 | 10 |
| STEM | none | -- | 1 | 0 | 10 |
| ANOVA | none | 1.5 | 0 | 0 | 0 |
| | none | 2 | 0 | 0 | 0 |
| | genes | 1.5 | 0 | 0 | 0 |
| | genes | 2 | 0 | 0 | 0 |

**Supplementary Materials:**

**Tables:**

**Table 1s: Biological and technical replicates of the microarrays used in this work.**

This table shows the biological replicates of the microarrays obtained from different rats exposed to hypoxia for 0, 1, 3, 7, 14, or 30 days. The number of technical replicates for each biological replicate is shown in the parenthesis beside the name of the biological replicate. There are 30 microarrays in total in this dataset.

| Day 0 | Day 1 | Day 3 | Day 7 | Day 14 | Day 30 |
|---|---|---|---|---|---|
| Hyp0D33 (5) | Hyp1D36 (1) | Hyp3D19 (1) | Hyp7D8 (1) | Hyp14D40 (1) | Hyp30D28 (2) |
| Hyp0D34 (4) | Hyp1D37 (1) | Hyp3D20 (1) | Hyp7D21 (1) | Hyp14D47 (3) | |
| Hyp0D35 (5) | Hyp1D38 (1) | | Hyp7D7 (2) | Hyp14D39 (1) | |

**Table 2s: The strategy for selecting microarrays for the simulated study in this work.**

We randomly selected 1 biological replicate of the microarrays for each time point of 1, 3, 7, and 30 days. Then for each chosen biological replicate, we randomly selected 1 technical replicate if multiple such replicates are available. For 0 and 14 days, we kept all the biological replicates; but for each biological replicate, we randomly selected 1 technical replicate if multiple such replicates were available. There are 10 microarrays in total in the simulated data.

| Day 0 | Day 1 | Day 3 | Day 7 | Day 14 | Day 30 |
|---|---|---|---|---|---|
| Hyp0D33 (r. s. 1*) | Hyp1D (r. s. 1) | Hyp3D (r. s. 1) | Hyp7D (r. s. 1) | Hyp14D40 (1) | Hyp30D28 (r. s. 1) |
| Hyp0D34 (r. s. 1) | | | | Hyp14D47 (r. s. 1) | |
| Hyp0D35 (r. s. 1) | | | | Hyp14D39 (1) | |

* r.s.1 is an abbreviation for 'randomly selected 1 array.'

**Table 3s. Pearson correlation of temporal expression patterns of the genes estimated by LRSA or STEM with those validated using RT-PCRs.**

Pearson correlation coefficients in this table were calculated as follow: first we calculated the relative expected expression values $\hat{f}_{i,0}(t)$ for any gene *i* with respect to its controls (0 day), and $\hat{f}_{i,0}(t) = 2 \wedge \left( \hat{f}_i(t) - \hat{f}_{i,0}(t=0) \right)$, where $\hat{f}_i(t)$ represents the log2-based expected expression values for gene *i* estimated by LRSA or STEM. Then we calculated the relative temporal expression values $E_{i,0}(t)$ of gene *i* measured using RT-PCRs. If we let $E_i(t)$ denote temporal expression values of gene *i* measured using RT-PCRs, then $E_{i,0}(t) = E_i(t)/E_i(t=0)$. Finally, the Pearson correlation coefficient was calculated to obtain the correlation of $\hat{f}_{i,0}(t)$ and $E_{i,0}(t)$ for gene *i*.

|      | Gene1 | Gene2 | Gene3 | Gene4 | Gene5 | Gene6 |
|------|-------|-------|-------|-------|-------|-------|
| LRSA | 0.89  | 0.79  | 0.87  | 0.95  | 0.64  | 0.59  |
| STEM | --    | 0.79  | 0.60  | 0.93  | --    | --    |

**Methods:**

**Microarray protocol:**

A CodeLink Bioarray experiment involves the following steps. Total RNAs are first prepared from a biological sample. Then a set of bacterial mRNAs of known concentrations (which are provided by the manufacturers and have complementary sequences to the positive control probes on the Bioarrays) are spiked in as positive controls. The mixed mRNAs are reverse transcribed into cDNAs and amplified into cRNAs, using *in vitro* transcription. The cRNAs are labeled with a fluorescent dye and hybridized to a CodeLink Bioarray presynthesized with probes. Finally the array is washed and scanned. The intensity value (signal) of the fluorescent dye detected from each probe on the array is proportional to the abundance of the targeting transcript in the sample of interest. The quality of these processes can be monitored by detecting the signal of the positive control probes on the Bioarrays.

**Microarray data preprocessing:**

Three rats were initially used for each experimental point. However, due to quality issues, several microarrays were eliminated from further data processing. The detailed information on the microarrays finally used in the downstream analyses can be found at the Gene Expression Omnibus (GEO) database with GEO serial accession number GSE8705 (http://www.ncbi.nlm.nih.gov/geo).

Before data preprocessing, data for all microarrays were summarized into data matrices. In each data matrix, genes (or probes) were arranged in rows; microarrays were in columns; data from the rats exposed to hypoxia for 0 day was used as controls.

To remove unwanted noise in the data, we preprocessed the microarray data using a procedure previously described by us in (Wu, et al., 2005). Briefly, we first removed the probes whose intensity values are missing in more than 5% of the microarrays. For the remaining probes, we used the *k-NNimpute* algorithm (Troyanskaya, et al., 2001) to fill in the intensity values which are missing in these probes. Then the processed microarray data was log2-transformed, and normalized using a statistical method, *CyclicLoess*, to minimize unwanted noise in the data (Wu, et al., 2005). We have shown previously that *CyclicLoess* is one of the most effective normalization methods for reducing intensity-dependent dye effects in CodeLink Bioarray data (Wu, et al., 2005). Finally, probes with very low signals were processed using a procedure recommended by the manufacturer of the arrays (Wu, et al., 2005).

Since there were both biological and technical replicates in the hypoxia dataset, we consolidated the data by merging (i.e., averaging) microarray data on multiple technical replicates obtained from the same rat into one single data vector; that is, let $s_{i,m}$ denote the intensity value of any gene $i$ on technical replicate $m$ and $1 \leq m \leq M$, then the intensity value $s_i$ of the same gene $i$ in the merged data can be obtained as:

$$s_i = \left( \sum_{m=1}^{M} s_{i,m} \right) / M .$$

The consolidated data was used for the LRSA analysis.

**Estimating simultaneous confidence intervals using simulated data:**

Since there was only one biological replicate for several time points (1, 3, 7, and 30 days) in the simulated data, the approximate simultaneous confidence intervals $I_i(t)$ cannot be estimated straightforwardly and accurately for the expected expression values $\hat{f}_i(t)$ at these time points (denoted as $T_{r1}$). In this work, we used a heuristic procedure to estimate $I_i(t)$ in these data. First, we used the method due to Faraway and Sun (Faraway and Sun, 1995) to estimate an approximate simultaneous confidence intervals $I_i(t)$ for $\hat{f}_i(t)$ for all *t*. Since there are multiple (3) biological replicates for time points 0 and 14 days, the simultaneous confidence intervals estimated for these time points, $I_i(t=T_{rm})$, where $T_{rm} = \{0,14\}$ should be accurate. Therefore we used $I_i(t=T_{rm})$ to estimate $I_i(t=T_{r1})$. To be conservative, we used the wider confidence intervals of these two time points $T_{rm}$ to estimate the simultaneous confidence intervals $I_i(t=T_u)$ for the other time points, where $T_u = (0,30]$ and $T_u \neq 14$, and thus,

$$I_i(t=T_u) = \left(\hat{f}_i(t) \pm \max\left(c_i \hat{\sigma}_i(t=0)\|l(t=0)\|, c_i \hat{\sigma}_i(t=14)\|l(t=14)\|\right)\right).$$